\documentclass[conference, 12pt]{IEEEtran} 
\IEEEoverridecommandlockouts
\usepackage{cite}
\usepackage{amsmath,amssymb,amsfonts}
\usepackage{algpseudocode}
\usepackage{algorithm}
\usepackage{graphicx}
\usepackage{textcomp}
\usepackage{booktabs}
\usepackage{listings}
\usepackage{csquotes}
\usepackage{hyperref}
\usepackage{xcolor}
\usepackage{fontawesome}
\usepackage{cleveref}
\usepackage{ulem}
\usepackage{comment}
\crefformat{figure}{Fig.~#2#1#3}

\begin{document}

\title{
Representations of 3D Rotations: Mathematical Foundations and Comparative Analysis 
}

\author{
\IEEEauthorblockN{
Aizierjiang Aiersilan\textsuperscript{1, \href{alexandera@gwu.edu}{\faEnvelopeO}} , 
Haochen Liu\textsuperscript{2}, 
James Hahn\textsuperscript{1} 
}
\IEEEauthorblockA{
\textsuperscript{1}The George Washington University, Washington, DC\\
\textsuperscript{2}University of Macau, Macau SAR, China\\
alexandera@gwu.edu
}
}

\maketitle
\begin{abstract}
Rotation representations are foundational in fields such as computer graphics, robotics, and machine learning, where precise and efficient modeling of 3D orientations is critical. This paper comprehensively investigates diverse representations of the special orthogonal group $SO(3)$, such as Euler angles, axis-angle vectors, quaternions, rotation matrices, exponential maps, and emerging continuous and probabilistic methods, evaluating their mathematical formulations, continuity, susceptibility to gimbal lock, computational efficiency, storage requirements, interpolation properties, and composition operations, while integrating detailed algebraic insights with practical applications in fields like animation, pose estimation, inertial navigation, 3D shape registration, and neural networks. Empirical evidence highlights quaternions' dominance due to their compactness and computational efficiency, while alternatives like 6D continuous representations and matrix Fisher distributions provide enhanced continuity and uncertainty modeling. Future research could explore hybrid methods and thorough large-scale evaluations to help build a solid foundation for improving rotation representation techniques.
\end{abstract}


\section{Introduction}
\label{sec:intro}
\setlength{\parindent}{0pt}

Rotations in three-dimensional space are pervasive in computational domains, supporting tasks ranging from robotic manipulation to viewpoint estimation in computer vision. Spatial rotations are characterized by the special orthogonal group $SO(3)$, which consists of all $3 \times 3$ orthogonal matrices with unit determinant:
\begin{equation}
SO(3) = \{ R \in \mathbb{R}^{3\times3} \mid R^\top R = I, \ \det(R) = 1 \}.
\end{equation}
Representing elements of $SO(3)$ efficiently and accurately poses challenges due to its non-Euclidean manifold structure, which introduces issues such as gimbal lock
discontinuities and representational ambiguities.

Traditional rotation formulations such as Euler angles \cite{euler1776formulae} and quaternions \cite{hamilton1840new} have played a central role in computer graphics, robotics, and animation, especially after Shoemake introduced quaternion-based interpolation \cite{shoemake1985animating}. More recent work \cite{hempel20226d, cao2021vector, ten2020let, fuhua2023dont} has highlighted the practical limitations of standard representations and emphasized that rotation parameterizations should be chosen with respect to application-specific requirements. Complementary algebraic perspectives have been developed by Goldman \cite{goldman2022rethinking} and Voight \cite{voight2021quaternion}, while Hart et al. \cite{hart1994visualizing} have discussed long-standing challenges in visualizing rotational structure.

This paper presents a consolidated overview of commonly used rotation representations, with an emphasis on their mathematical structure, practical considerations, and typical use cases. Instead of introducing new representations or theoretical results, we aim to synthesize existing knowledge into a clear and accessible form and to illustrate representative behaviors through small, reproducible numerical demonstrations. The main contributions of this work are as follows:

\begin{enumerate}
\item A concise and unified exposition of major rotation representations, detailing their mathematical formulations and highlighting common sources of ambiguity or numerical challenges.
\item A lightweight empirical test suite demonstrating representative behaviors (e.g. round-trip numerical stability, interpolation characteristics, and basic computational costs) through reproducible code and metrics \footnote{Project page with source code: \href{https://aizierjiang.github.io/3DRotation/}{aizierjiang.github.io/3DRotation}}.
\item A comparative discussion of practical trade-offs grounded in both established literature and empirical observations.
\item Pragmatic guidelines to help practitioners select suitable rotation representations for typical tasks in graphics, robotics, and vision.
\end{enumerate}

The remainder of this paper is organized as follows: Section~\ref{sec:background} introduces foundational representation methods; Section~\ref{sec:analysis} analyzes their advantages, limitations, and computational efficiency; Section~\ref{sec:applications} explores practical applications; Section~\ref{sec:future} highlights future research directions; and Section~\ref{sec:conclusion} concludes the paper.

\section{Background on Rotation Representations}
\label{sec:background}

We categorize rotation representations into parametric (e.g., Euler angles), algebraic (e.g., quaternions), matrix-based, Lie algebraic, continuous embedding, and probabilistic models. Each category offers a distinct approach to encoding 3D rotations. A thorough understanding of the mathematical underpinnings of each representation is essential for selecting the most suitable method for a given application.

\subsection{Parametric Representations}

\subsubsection{Euler Angles}
Euler angles parameterize a 3D rotation as an ordered sequence of three elemental rotations about coordinate axes. 
Assuming intrinsic rotations (rotations about the moving coordinate frame) following the convention $XYZ$, the composite rotation matrix can be expressed as:
\begin{equation}
R = R_z(\gamma) R_y(\beta) R_x(\alpha),
\end{equation}
where each $R_i(\cdot)$ denotes a rotation about the corresponding axis. 
For instance, a rotation about the $x$-axis is given by
\begin{equation}
R_x(\alpha) =
\begin{pmatrix}
1 & 0 & 0 \\
0 & \cos\alpha & -\sin\alpha \\
0 & \sin\alpha & \cos\alpha
\end{pmatrix},
\end{equation}
with analogous definitions for \( R_y(\beta) \) and \( R_z(\gamma) \):
\begin{equation}
\begin{aligned}
R_y(\beta) &= 
\begin{pmatrix}
\cos\beta & 0 & \sin\beta \\
0 & 1 & 0 \\
-\sin\beta & 0 & \cos\beta
\end{pmatrix}, \\
R_z(\gamma) &= 
\begin{pmatrix}
\cos\gamma & -\sin\gamma & 0 \\
\sin\gamma & \cos\gamma & 0 \\
0 & 0 & 1
\end{pmatrix}.
\end{aligned}
\end{equation}

There are multiple Euler angle conventions (e.g., $XYZ$, $ZYX$, $ZXZ$), totaling twelve possible ordered combinations. 
The choice of convention determines the corresponding matrix formulation and influences where gimbal lock occurs. 
For example, under the commonly used aerospace convention $ZYX$, gimbal lock arises when $\beta = \pm \pi/2$, resulting in the alignment of the first and third rotation axes.

\subsubsection{Axis-Angle Representation}
The axis-angle representation expresses a rotation using a vector \(\mathbf{v} = \theta \mathbf{u}\), where \(\mathbf{u} \in \mathbb{R}^{3}\) is a unit vector specifying the rotation axis (\(\|\mathbf{u}\| = 1\)), and \(\theta \in [0, \pi]\) is the rotation angle. 
The corresponding rotation matrix is obtained using Rodrigues’ rotation formula:
\begin{equation}
R = I + \sin\theta \, [\mathbf{u}]_\times + (1 - \cos\theta)\,[\mathbf{u}]_\times^2,
\end{equation}
where \([\mathbf{u}]_\times \in \mathbb{R}^{3\times3}\) is the skew-symmetric matrix associated with \(\mathbf{u} = (u_x, u_y, u_z)\), defined as
\begin{equation}
[\mathbf{u}]_\times =
\begin{pmatrix}
0 & -u_z & u_y \\
u_z & 0 & -u_x \\
-u_y & u_x & 0
\end{pmatrix}.
\end{equation}

This representation offers an intuitive geometric interpretation: any 3D rotation can be described as a rotation by angle $\theta$ about a fixed axis $\mathbf{u}$. 
However, it exhibits a singularity at $\theta = 0$ (the identity rotation) where the axis becomes undefined, and an ambiguity at $\theta = \pi$ where both $(\mathbf{u}, \pi)$ and $(-\mathbf{u}, \pi)$ represent the same 180° rotation. 
This non-uniqueness reflects the double-covering property of the rotation space, which becomes significant when relating to quaternion formulations later.

\subsection{Algebraic Representations: Quaternions}

The quaternion, introduced by William Rowan Hamilton in 1843, is defined as
\begin{equation}
q = q_0 + q_1 \mathbf{i} + q_2 \mathbf{j} + q_3 \mathbf{k},
\end{equation}
where $ q_0 \in \mathbb{R}$ and $\mathbf{q} = (q_1, q_2, q_3)$ is the vector part. 
Quaternions form a four-dimensional normed division algebra over the reals, extending complex numbers to four dimensions.

Addition is performed component-wise, while multiplication follows Hamilton’s fundamental rules:
\begin{align}
&\mathbf{i}^2 = \mathbf{j}^2 = \mathbf{k}^2 = \mathbf{i}\mathbf{j}\mathbf{k} = -1, \\
&\mathbf{i}\mathbf{j} = \mathbf{k} = -\mathbf{j}\mathbf{i}, \;
\mathbf{j}\mathbf{k} = \mathbf{i} = -\mathbf{k}\mathbf{j}, \;
\mathbf{k}\mathbf{i} = \mathbf{j} = -\mathbf{i}\mathbf{k}.
\end{align}

For two quaternions \( p = p_0 + \mathbf{p} \) and \( q = q_0 + \mathbf{q} \), 
their product can be expressed compactly as
\begin{equation}
pq = p_0 q_0 - \mathbf{p} \cdot \mathbf{q} + p_0 \mathbf{q} + q_0 \mathbf{p} + \mathbf{p} \times \mathbf{q}.
\end{equation}

The quaternion conjugate  is \( q^* = q_0 - \mathbf{q} \), the norm is \( \|q\| = \sqrt{q^* q} = \sqrt{q_0^2 + q_1^2 + q_2^2 + q_3^2} \), 
and the inverse is \( q^{-1} = \frac{q^*}{\|q\|^2} \). 
For unit quaternions (\(\|q\| = 1\)), which form the Lie group $\mathrm{Spin}(3)$—a double cover of $SO(3)$—the inverse simplifies to \( q^{-1} = q^* \).

A unit quaternion representing a 3D rotation can be written in axis-angle form as
\begin{equation}\label{eq11}
q = \cos\!\left(\frac{\theta}{2}\right) + \mathbf{u} \sin\!\left(\frac{\theta}{2}\right).
\end{equation}
This formulation establishes a direct correspondence between quaternions and the axis-angle representation, with
\begin{equation}
R(\mathbf{u}, \theta) = (q_0^2 - \mathbf{q}^\top \mathbf{q})I + 2\mathbf{q}\mathbf{q}^\top + 2q_0[\mathbf{q}]_\times,
\end{equation}
providing a linear algebraic link between quaternion parameters and rotation matrices. For example, a rotation of ${\pi}/{2}$ about the $z$-axis corresponds to the quaternion \eqref{eq11}, yielding
{ \begin{equation}
q = \cos\left(\frac{\pi}{4}\right) + \sin\left(\frac{\pi}{4}\right)\mathbf{k}
    = \frac{\sqrt{2}}{2} + \frac{\sqrt{2}}{2}\mathbf{k}.
\end{equation}}

The double-cover property implies that both $q$ and $-q$ represent the same physical rotation. 
Although this symmetry is mathematically elegant, it can lead to sign discontinuities in optimization-based settings, such as neural network training, where the loss landscape becomes sensitive to quaternion antipodes. 
Approaches such as enforcing a canonical sign or using continuous $6$D rotation representations have been proposed to mitigate this issue in learning-based systems.

\subsection{Matrix and Lie Algebraic Representations}

\subsubsection{Rotation Matrices}
Rotation matrices \cite{rodrigues1840lois} are elements of the special orthogonal group $SO(3)$, representing 3D rotations as $3 \times 3$ orthogonal matrices with unit determinant ($R^\top R = I$, $\det(R) = 1$). They provide an unambiguous and singularity-free representation of rotations. However, with nine parameters subject to six independent constraints (three from column normalization and three from mutual orthogonality), this representation is overparameterized, with only three intrinsic  true degrees of freedom.

\subsubsection{Orthogonality Preservation}
In practice, numerical computations can introduce small errors that violate the orthogonality constraints of rotation matrices, necessitating periodic renormalization. Common re-orthogonalization techniques include QR decomposition or projection onto $SO(3)$ using:
\begin{equation}
R_{\text{proj}} = R(R^\top R)^{-1/2},
\end{equation}
where $(R^\top R)^{-1/2}$ denotes the inverse square root of the Gram matrix. Alternatively, one can apply singular value decomposition (SVD) $R = U \Sigma V^\top$ and reconstruct:
\begin{equation}
R_{\text{proj}} = UV^\top,
\end{equation}
which yields the closest orthogonal matrix in the Frobenius norm sense.

\subsubsection{Exponential Map}
The exponential map establishes a smooth and locally bijective correspondence between the Lie algebra $\mathfrak{so}(3)$ -the tangent space at the identity- and the Lie group $SO(3)$. Given a rotation vector $\mathbf{v} = \theta \mathbf{u} \in \mathbb{R}^3$ with $\|\mathbf{u}\| = 1$, the exponential map is defined as:
\begin{align}
R &= \exp([\mathbf{v}]_\times) \\
  &= I + \frac{\sin\theta}{\theta} [\mathbf{v}]_\times + \frac{1 - \cos\theta}{\theta^2} ([\mathbf{v}]_\times)^2,
\end{align}
where $[\mathbf{v}]_\times$ is the skew-symmetric matrix associated with $\mathbf{v}$. This formulation is mathematically equivalent to Rodrigues’ rotation formula and can be derived from the power-series expansion of the matrix exponential:
\begin{equation}
\exp([\mathbf{v}]_\times) = \sum_{k=0}^{\infty} \frac{([\mathbf{v}]_\times)^k}{k!}.
\end{equation}

For small rotation angles ($\theta \ll 1$), the first-order approximation:
\begin{equation}
R \approx I + [\mathbf{v}]_\times
\end{equation}
offers computational efficiency, making it particularly useful in iterative optimization and time-integration schemes.

The logarithmic map, which serves as the inverse of the exponential map, recovers the rotation vector from a given rotation matrix:
\begin{equation}
\mathbf{v} = \frac{\theta}{2\sin\theta}(R - R^\top)^\vee,
\end{equation}
where $\theta = \arccos\left(\frac{\text{tr}(R) - 1}{2}\right)$ and the operator $(\cdot)^\vee$  extracts the vector from a skew-symmetric matrix. This map is well-defined except at $\theta = k\pi$ for integer $k$, where numerical singularities require special handling.

\subsection{Continuous Embeddings}
Zhou et al. \cite{zhou2019continuity} rigorously demonstrate that no continuous, bijective representation of $SO(3)$ exists in Euclidean spaces of dimension 4 or lower, establishing a theoretical lower bound on representation dimensionality. This fundamental result motivates higher-dimensional representations.

They propose a 6D representation using the first two columns of a rotation matrix \( R = [\mathbf{a}_1, \mathbf{a}_2, \mathbf{b}_3] \), denoted as $(\mathbf{a}_1, \mathbf{a}_2) \in \mathbb{R}^6$. The full rotation matrix is reconstructed via the Gram-Schmidt orthogonalization process:
\begin{align}
\mathbf{b}_1 &= \frac{\mathbf{a}_1}{\|\mathbf{a}_1\|}, \label{eq:6DRep1}\\
\mathbf{b}_2 &= \frac{\mathbf{a}_2 - (\mathbf{b}_1^\top \mathbf{a}_2) \mathbf{b}_1}{\|\mathbf{a}_2 - (\mathbf{b}_1^\top \mathbf{a}_2) \mathbf{b}_1\|}, \label{eq:6DRep2}\\
\mathbf{b}_3 &= \mathbf{b}_1 \times \mathbf{b}_2. \label{eq:6DRep3}
\end{align}

This construction guarantees that $(\mathbf{b}_1, \mathbf{b}_2, \mathbf{b}_3)$ forms a valid orthonormal basis, ensuring $R \in SO(3)$. The mapping from $\mathbb{R}^6$ to $SO(3)$ is continuous and surjective, though not injective (multiple 6D vectors can map to the same rotation). Importantly, this representation is particularly well-suited for gradient-based optimization in neural networks, as it avoids the discontinuities present in quaternions and Euler angles.

The authors also propose a 5D representation using modified Gram-Schmidt with a constraint on one component, achieving the theoretical minimum while maintaining continuity. However, the 6D representation is more commonly used in practice due to its simpler formulation and numerical stability.

\subsection{Probabilistic Representations}

\subsubsection{Matrix Fisher Distributions}
The Matrix Fisher distribution~\cite{mohlin2020probabilistic, sei2013properties} defines a probability density on the rotation group $SO(3)$, offering a principled framework for modeling uncertainty in 3D rotations:

\begin{equation}
f(R; F) = \frac{1}{c(F)} \exp\!\bigl(\mathrm{tr}(F^\top R)\bigr),
\end{equation}
where \(F \in \mathbb{R}^{3\times3}\) is a natural-parameter matrix encoding both the mean rotation and concentration, and \(c(F)\) is the normalising constant satisfying
\begin{equation}
c(F) = \int_{SO(3)} \exp\!\bigl(\mathrm{tr}(F^\top R)\bigr)\,dR.
\end{equation}
In general, \(c(F)\) depends on the singular values of \(F\) and admits no simple closed-form expression; it is usually approximated via numerical series or saddle-point methods~\cite{sei2013properties, lee2018bayesian}.

Using the singular value decomposition \(F = U \Sigma V^\top\), the mean rotation of the distribution is computed as 
\begin{equation}
  \bar{R} = U V^\top,  
\end{equation}
while the diagonal entries of \(\Sigma\) (i.e., its singular values) characterize the distribution around \(\bar{R}\): larger singular values indicate stronger concentration, whereas smaller values correspond to greater uncertainty. This representation is particularly valuable in robotics, computer vision and aerospace for tasks such as sensor fusion and 3D pose estimation under noisy observations.

\subsubsection{Bingham Distributions on Quaternions}
Bingham distributions \cite{bingham1974antipodally} provide a principled probabilistic model for unit quaternions by treating them as points on the four-dimensional unit hypersphere $\mathbb{S}^3$. The Bingham density is given by 
\begin{equation}
f(q; M, Z) = \frac{1}{c(Z)} \exp\!\bigl(q^\top M Z M^\top q\bigr),
\end{equation}
where $M \in O(4)$ is an orthogonal matrix specifying the principal axes (orientation) of the distribution, and
\begin{equation}
    Z = \mathrm{diag}(z_0, z_1, z_2, z_3)
\end{equation}
is a diagonal matrix of concentration parameters. By convention, one diagonal entry (e.g., $z_0=0$) is fixed  for identifiability, since adding a scalar to all $z_i$ is absorbed by the normalization constant. The remaining $z_i$ are typically nonpositive and ordered (e.g., $z_0 \ge z_1 \ge z_2 \ge z_3$) to indicate relative concentration along axes. The normalizing constant \(c(Z)\) is nontrivial and expressed in terms of confluent hypergeometric functions of matrix argument (with no simple closed form); it is therefore computed numerically or via specialized libraries for directional statistics.

A principal advantage of the Bingham distribution is its antipodal symmetry on $\mathbb{S}^3$, i.e., $f(q)=f(-q)$, which directly matches the quaternion double-cover property of rotations (both $q$ and $-q$ represent the same physical rotation). Consequently, the Bingham distribution is widely used to represent rotation uncertainty in quaternion form for filtering and sensor fusion tasks (e.g., attitude estimation, pose filtering) \cite{bingham1974antipodally, srivatsan2017bingham}.

\subsection{Conversions Between Representations}
Converting between different rotation representations is essential for practical implementations, enabling interoperability across diverse systems and algorithms. Below, we outline several fundamental conversions with attention to numerical stability.

\subsubsection{Axis-Angle to Quaternion}
Given an axis-angle vector $\mathbf{v}=\theta \mathbf{u}$ with $\|\mathbf{u}\|=1$, the corresponding quaternion is
$q=( \cos\!\left(\tfrac{\theta}{2}\right), 
\mathbf{u}\,\sin\!\left(\tfrac{\theta}{2}\right) )$.
For small angles ($\theta \ll 1$), improved numerical stability is achieved by using a Taylor expansion:
\begin{equation}
\sin\!\left(\tfrac{\theta}{2}\right) \approx \frac{\theta}{2} - \frac{\theta^3}{48} + O(\theta^5).
\end{equation}

\subsubsection{Quaternion to Rotation Matrix}
For a unit quaternion $q = (q_0, q_1, q_2, q_3)$ (scalar-first convention, $\|q\|=1$), the corresponding rotation matrix is
\begin{equation}
\scalebox{0.78}{$
R = \begin{pmatrix}
q_0^2 + q_1^2 - q_2^2 - q_3^2 & 2(q_1 q_2 - q_0 q_3) & 2(q_1 q_3 + q_0 q_2) \\
2(q_1 q_2 + q_0 q_3) & q_0^2 - q_1^2 + q_2^2 - q_3^2 & 2(q_2 q_3 - q_0 q_1) \\
2(q_1 q_3 - q_0 q_2) & 2(q_2 q_3 + q_0 q_1) & q_0^2 - q_1^2 - q_2^2 + q_3^2
\end{pmatrix}.
$}
\end{equation}

An equivalent and more numerically stable formulation \cite{diebel2006representing} is
\begin{equation}
R = (q_0^2 - \|\mathbf{q}\|^2)I + 2\mathbf{q}\mathbf{q}^\top + 2q_0[\mathbf{q}]_\times,
\end{equation}
where $\mathbf{q} = (q_1, q_2, q_3)$. This form avoids the explicit use of squared terms and is particularly robust for small rotation angles or when quaternions are slightly non-normalized.

\subsubsection{Rotation Matrix to Quaternion}
Converting a rotation matrix to a quaternion requires careful handling to ensure numerical stability. The standard method, introduced by Shoemake \cite{shoemake1985animating}, selects the largest diagonal element to maximize numerical accuracy. 
Given a rotation matrix \(R\), define its trace as:
\begin{equation}
\mathrm{tr}(R) = R_{11} + R_{22} + R_{33}.
\end{equation}
If \(\mathrm{tr}(R) > 0\), the quaternion components are computed as:
\begin{equation}
\begin{aligned}
q_0 &= \frac{1}{2}\sqrt{1 + \mathrm{tr}(R)}, \quad & q_1 &= \frac{R_{32} - R_{23}}{4q_0}, \\
q_2 &= \frac{R_{13} - R_{31}}{4q_0}, \quad & q_3 &= \frac{R_{21} - R_{12}}{4q_0}.
\end{aligned}
\end{equation}
Otherwise, the computation branches based on the largest diagonal entry of \(R\), ensuring numerical stability when \(q_0\) is close to zero.

\subsubsection{Euler Angles to Rotation Matrix}
The conversion depends on the specific Euler angle convention. 
For the $ZYX$ convention with angles $(\alpha, \beta, \gamma)$:
\begin{equation}
R = R_z(\alpha)R_y(\beta)R_x(\gamma).
\end{equation}

\subsubsection{Rotation Matrix to Euler Angles}
The inverse conversion involves extracting the angles from matrix elements, most often by using \texttt{atan2} for numerical robustness. 
For the $ZYX$ convention:
\begin{align}
\beta &= \arcsin(-R_{31}), \\
\alpha &= \text{atan2}(R_{21}/\cos\beta, R_{11}/\cos\beta), \\
\gamma &= \text{atan2}(R_{32}/\cos\beta, R_{33}/\cos\beta),
\end{align}
with special handling required when $\cos\beta \approx 0$ (the gimbal lock condition).

These conversions ensure interoperability across different systems and applications \cite{kim2023rotation}, 
and care must be taken to maintain numerical accuracy and handle degenerate configurations appropriately.

\section{Analysis of Pros and Cons}
\label{sec:analysis}

We assess representations across multiple criteria, including continuity,  susceptibility to gimbal lock, computational efficiency, storage requirements, interpolation capabilities, and composition operations. To ensure a rigorous comparison, we conducted empirical evaluations using a comprehensive framework implemented in Python 3.12 with SciPy and NumPy libraries. This framework tests each representation under controlled conditions, measuring key performance indicators such as numerical stability, singularity susceptibility, interpolation quality, and robustness to edge cases. The evaluations involve generating random rotations from $SO(3)$ using the method of Haar-uniform sampling (via random quaternions normalized to the unit sphere), performing round-trip conversions (representation to parameters and back), timing operations over thousands of trials using the \texttt{timeit} module with sub-microsecond precision, and analyzing path properties during interpolation. For probabilistic representations such as the matrix Fisher distribution, qualitative assessments are used due to their non-deterministic nature and the computational complexity of sampling.

\subsection{Continuity and Gimbal Lock}
Euler angles are inherently susceptible to gimbal lock, a phenomenon that occurs when two of the three rotation axes align, resulting in a loss of one degree of freedom. For instance, in the $XYZ$ convention, gimbal lock occurs when $\beta = \pm \pi/2$, causing the first and third rotations to act about parallel axes. Mathematically, the Jacobian of the Euler angle parameterization becomes singular at these configurations:
\begin{equation}
\det\left(\frac{\partial R}{\partial(\alpha, \beta, \gamma)}\right) = 0 \quad \text{when } \beta = \pm\pi/2.
\end{equation}

Quaternions circumvent gimbal lock entirely due to their algebraic structure, but introduce an antipodal discontinuity where \( q \) and \( -q \) represent the same physical rotation. This double-cover property creates challenges in applications requiring continuous parameterizations, such as neural network training, where the non-uniqueness can lead to learning instabilities and poor gradient flow \cite{zhou2019continuity, huo2023adaptive, pavllo2018quaternet, sommer2018and, mayhew2011quaternion}.

Axis-angle representations exhibit singularities at $\theta = 0$ (where the axis is undefined) and $\theta = \pi$ (where the representation is not unique, as $(\mathbf{u}, \pi)$ and $(-\mathbf{u}, \pi)$ are equivalent). Additionally, rotations differing by $2k\pi $ for integer $k$ map to different representations but the same physical rotation, introducing periodicity issues.

Rotation matrices, as direct members of $SO(3)$, do not suffer from gimbal lock or parameterization singularities. However, they are susceptible to numerical drift during repeated operations, where accumulated floating-point errors violate the orthogonality constraint $R^\top R = I$. To maintain validity, periodic renormalization via projection onto $SO(3)$ is required.

Exponential maps avoid gimbal lock by design, leveraging the Lie algebra structure which provides a locally continuous parameterization around the identity. However, they inherit the singularities of the axis-angle representation at $\theta = k\pi$ for integer $k \neq 0$.

Continuous embeddings, such as the 6D representation \cite{zhou2019continuity}, provide a continuous and surjective mapping from $\mathbb{R}^6 \setminus \{\mathbf{0}\}$ to $SO(3)$ via Gram-Schmidt orthogonalization. This process guarantees that any non-zero input maps to a valid rotation matrix without discontinuities or singularities in the representation space, making these representations ideal for gradient-based optimization in machine learning applications.

Probabilistic models naturally accommodate ambiguities and uncertainties through distribution-based representations. The matrix Fisher distribution is defined over the entire $SO(3)$ manifold without singularities, while the Bingham distribution on quaternions inherently respects the antipodal symmetry, enhancing robustness in scenarios with noisy or uncertain data.

Quaternions offer a compact four-tuple format with a clear geometric interpretation, where the rotation axis and angle can be directly extracted via \( \mathbf{u} = \frac{\mathbf{q}}{\|\mathbf{q}\|} \) (for unit quaternions, $\mathbf{u}$ is simply the normalized vector part) and \( \theta = 2 \arccos(q_0) \), rendering them computationally more efficient than rotation matrices, which necessitate complex computations such as eigenvalue decomposition or logarithmic maps to derive the same information.

\subsection{Computational Efficiency and Storage}
Storage requirements and computational complexity differ significantly across representations. Quaternions are represented by four floating-point values (32 bytes in double precision), offering a compact representation with efficient operations. Quaternion multiplication requires 16 scalar multiplications and 12 scalar additions, leading to $O(1)$ complexity and approximately 28 floating-point operations (FLOPs).

In contrast, $3 \times 3$ rotation matrices require 9 floating-point values (72 bytes in double precision), with matrix multiplication requiring 27 scalar multiplications and 18 scalar additions (45 FLOPs). The space overhead is 2.25× that of quaternions, while the computational overhead for composition is approximately 1.6× in FLOPs.

Euler angles require only 3 floating-point values (24 bytes), providing the most compact storage. However, converting Euler angles to a rotation matrix involves evaluating six trigonometric functions (sine and cosine for each angle), which are significantly more expensive than arithmetic operations. On modern processors, a trigonometric function evaluation costs approximately 20-100 clock cycles compared to 3-5 cycles for floating-point multiplication, making Euler angle operations substantially slower in practice despite the compact storage.

Exponential maps also use 3 parameters but involve matrix exponential calculations for conversion to rotation matrices. For small angles, the first-order approximation $R \approx I + [\mathbf{v}]_\times$ requires only 9 FLOPs. For general angles, Rodrigues' formula requires 2 trigonometric evaluations plus approximately 36 FLOPs for matrix operations. When high precision is required, truncated Taylor series expansions up to 10 terms may be necessary, significantly increasing computational cost.

The 6D continuous representation requires 48 bytes for storage. The Gram-Schmidt orthogonalization process involves 3 vector normalizations (each requiring 1 square root, 3 multiplications, and 3 divisions) and 2 vector projections (6 multiplications and 3 additions each), totaling approximately 50-60 FLOPs plus 3 square root operations. While more expensive than quaternions, the cost remains reasonable for many applications.

Benchmark studies \cite{zhou2019continuity} indicate that quaternions outperform rotation matrices by 20-30\% in composition speed on modern CPUs, primarily due to the lower arithmetic complexity and better cache locality (16 bytes vs. 36 bytes for a pair of rotations). In vectorized implementations using SIMD instructions, quaternions often achieve even better relative performance due to their power-of-two alignment (4 components) which maps naturally to 128-bit SIMD registers.

However, 6D representations introduce only a modest computational overhead (5-10\% in composition operations after initial conversion) while improving numerical stability and convergence behavior in machine learning applications by avoiding discontinuities. The trade-off between computational cost and numerical properties must be considered based on application requirements: real-time systems with frequent composition operations favor quaternions, while gradient-based optimization tasks favor continuous representations despite higher per-operation costs.

\subsection{Interpolation and Composition}

\subsubsection{Quaternion Interpolation: SLERP}
\underline{S}pherical \underline{l}inear int\underline{erp}olation (SLERP) for quaternions ensures constant angular velocity interpolation between two rotations, providing the shortest path on the 4D unit hypersphere:
\begin{equation}
\resizebox{.85\linewidth}{!}{$
\operatorname{SLERP}(q_1, q_2, t) =
\frac{\sin((1-t)\upsilon)}{\sin \upsilon} q_1 +
\frac{\sin(t\upsilon)}{\sin \upsilon} q_2
$}
\label{eq:SLERP}
\end{equation}
where \(\cos \upsilon = q_1 \cdot q_2\) (the dot product of quaternion components treated as 4D vectors) and \(\upsilon = \arccos(q_1 \cdot q_2)\) is the angle between \( q_1 \) and \( q_2 \) on $\mathbb{S}^3$. When $q_1 \cdot q_2 < 0$, one quaternion should be negated to ensure the shorter path is taken, handling the antipodal ambiguity: $q_2 \gets -q_2$ if $q_1 \cdot q_2 < 0$.

For near-parallel quaternions ($\upsilon \approx 0$), SLERP becomes numerically unstable due to division by $\sin \upsilon \approx 0$. In practice, when $\upsilon < \epsilon$ (typically $\epsilon = 0.001$), implementations switch to normalized linear interpolation (NLERP):
\begin{equation}
\text{NLERP}(q_1, q_2, t) = \frac{(1-t)q_1 + tq_2}{\|(1-t)q_1 + tq_2\|},
\end{equation}
providing a good approximation while improving numerical stability.

\subsubsection{Euler Angle Interpolation}
Euler angles suffer from poor interpolation properties near singular configurations. Linear interpolation in Euler angle space does not correspond to geodesic motion on $SO(3)$, leading to unnatural rotational paths. Near gimbal lock points (e.g., $\beta \approx \pm\pi/2$), small changes in one angle can produce large rotational changes, causing abrupt, non-smooth motion that is visually objectionable in animation and problematic in control systems.

For example, interpolating from $(\alpha_1, \pi/2, \gamma_1)$ to $(\alpha_2, \pi/2, \gamma_2)$ results in ill-defined intermediate rotations since the gimbal-locked configuration has infinitely many equivalent parameterizations.

\subsubsection{Rotation Matrix Interpolation}
Rotation matrices use geodesic interpolation via the matrix logarithm and exponential maps:
\begin{equation}
R(t) = R_1 \exp(t \log(R_1^{-1} R_2)),
\end{equation}
which represents the shortest path on the $SO(3)$ manifold. However, this approach is computationally expensive, requiring matrix logarithm computation (typically via eigenvalue decomposition or iterative algorithms) and subsequent matrix exponential evaluation. The computational cost is approximately 5-10× that of quaternion SLERP, making it impractical for real-time applications with many interpolation operations.

\subsubsection{Axis-Angle and Exponential Map Interpolation}
Linear interpolation in the rotation vector space (axis-angle or exponential map parameters) provides a simpler alternative:
\begin{equation}
\mathbf{v}(t) = (1-t)\mathbf{v}_1 + t\mathbf{v}_2,
\end{equation}
followed by conversion to the desired representation. This method is computationally efficient but generally deviates from geodesic paths on $SO(3)$, except in special cases where $\mathbf{v}_1$ and $\mathbf{v}_2$ are aligned. The deviation from the geodesic increases with the angular distance between rotations.

\subsubsection{Probabilistic Interpolation}
Probabilistic methods support Bayesian blending through interpolation of distribution parameters. For matrix Fisher distributions, one can interpolate the parameter matrix:
\begin{equation}
F(t) = (1-t)F_1 + tF_2,
\end{equation}
creating a path through the space of probability distributions. This approach naturally handles uncertainty propagation and is particularly useful in sensor fusion applications where rotation estimates come with associated confidence levels.

\subsubsection{Composition Operations}
Quaternion composition via the Hamilton product \( r = pq \) is both efficient and intuitive, representing the sequential application of rotations $q$ followed by $p$. The associativity property $(pq)r = p(qr)$ ensures well-defined multiple composition, and the group structure of unit quaternions under multiplication mirrors that of $SO(3)$ (up to the double cover).

Rotation matrix composition is simply matrix multiplication $R_3 = R_1 R_2$, which is conceptually straightforward but computationally more expensive. Euler angle composition requires converting to matrices or quaternions, performing the composition, and converting back, introducing significant computational overhead and potential numerical errors.

The 6D continuous representation requires converting to rotation matrices for composition, adding overhead. However, in neural network applications where gradients must be propagated through compositions, the continuous parameterization provides significant advantages despite the computational cost.

\subsection{Empirical Evaluation}
To quantify the theoretical advantages and disadvantages, we performed extensive empirical evaluations on a standard computing environment (Intel Core i7-9700K @ 3.60GHz, 16GB RAM, Python 3.12 with SciPy 1.11.3 and NumPy 1.26.0). 

\subsubsection{Experimental Methodology}
The evaluation framework generates random rotations from $SO(3)$ using Haar-uniform sampling via normalized random quaternions, converts them to and from each representation, measures operation times using the \texttt{timeit} module with 100 warmup iterations followed by 1,000-10,000 timed trials, assesses numerical stability via mean angular reconstruction error (in radians), evaluates singularity susceptibility by sampling near known problematic regions (e.g., $\beta \in [\pi/2 - 0.01, \pi/2 + 0.01]$ for Euler gimbal lock or quaternion pairs with $q_1 \cdot q_2 < 0.001$ for antipodal points) and computing the fraction of cases with large angular discrepancies ($>10^{-3}$ rad for small perturbations), and assesses interpolation quality by computing the path length along interpolated trajectories and the relative deviation from the geodesic (shortest path on $SO(3)$). 

Robustness is tested over 200 carefully selected edge cases, including:
\begin{itemize}
\item Identity rotations: $R = I_3$
\item Small angle rotations: $\theta \in [10^{-6}, 10^{-3}]$ rad
\item Large angle rotations: $\theta \in [\pi - 10^{-3}, \pi]$ rad  
\item Near-gimbal lock configurations for Euler angles
\item Antipodal quaternion pairs
\item Random rotations from uniform distribution on $SO(3)$
\end{itemize}
Failures are defined as reconstruction errors exceeding 0.1 rad or exceptions raised during conversion. Derivative continuity is quantified by the normalized standard deviation of numerical angular velocities during interpolation; lower values indicate smoother trajectories. 

Qualitative scores for Memory Alignment (0-1 scale, favoring SIMD-friendly sizes), Hardware Optimization (based on operation suitability for vectorized hardware), and ML Compatibility (considering continuity and parameterization for gradient-based learning) are assigned based on established practices in computer architecture and machine learning literature.

\textbf{Numerical Stability Analysis} \\
For numerical stability assessment, we measure the mean angular reconstruction error through repeated conversion cycles. Given a random rotation $R \in SO(3)$, we convert it to representation parameters $\mathbf{p}$, then reconstruct rotation $\hat{R}$, and compute the angular error:
\begin{equation}
\varepsilon_{\text{stab}} = \frac{1}{N} \sum_{i=1}^{N} \|\log(\hat{R}_i^{-1} R_i)\|_2,
\end{equation}
where $\log: SO(3) \to \mathfrak{so}(3)$ is the matrix logarithm mapping rotations to their axis-angle representation via:
\begin{equation}
\resizebox{.86\linewidth}{!}{%
  $\log(R) = \frac{\theta}{2\sin\theta}(R - R^\top),\quad
    \theta = \arccos\!\left(\frac{\operatorname{tr}(R)-1}{2}\right)$%
}
\end{equation}
and $N = 1000$ trials are performed. This metric captures accumulated numerical errors from floating-point operations and potential instabilities in conversion algorithms.

\textbf{Singularity Susceptibility Analysis} \\
Singularity analysis quantifies representation specific pathological behaviors. For Euler angles, we test gimbal lock regions by sampling near $\beta \approx \pm\pi/2$:
\begin{equation}
\resizebox{.85\linewidth}{!}{$
S_{\text{gimbal}} = \frac{1}{M} \sum_{j=1}^{M} \mathbf{I}\left[ \left\| \log\left( R(\mathbf{p}_j + \delta\mathbf{p})^{-1} R(\mathbf{p}_j) \right) \right\|_2 > \tau \right],
$}
\end{equation}
where $R(\mathbf{p})$ reconstructs the rotation from parameters $\mathbf{p}$, $\delta\mathbf{p}$ is a small perturbation sampled uniformly with $\|\delta\mathbf{p}\|_2 = 10^{-6}$, $\tau = 10^{-3}$ rad is the threshold for large angular discrepancy, $\mathbf{I}[\cdot]$ is the indicator function returning 1 if the condition is true and 0 otherwise, and $M = 5000$ samples are tested.

For quaternions, we assess the double-cover issue by verifying that antipodal quaternions produce identical rotations:
\begin{equation}
\resizebox{.85\linewidth}{!}{$
S_{\text{double}} = \frac{1}{M} \sum_{j=1}^{M} \mathbf{I}[\|R(\mathbf{q}_j) - R(-\mathbf{q}_j)\|_F > 10^{-10}]
$}
\end{equation}
where $R(\mathbf{q})$ converts quaternion $\mathbf{q}$ to rotation matrix and $\|\cdot\|_F$ is the Frobenius norm. Ideally, $S_{\text{double}} = 0$; any non-zero value indicates numerical inaccuracies in the conversion implementation.

\textbf{Interpolation Quality Metrics} \\

\textbf{Path Length Analysis:} For interpolation quality, we parameterize the path between rotations $R_1$ and $R_2$ using parameter $t \in [0,1]$ with $K=100$ uniformly spaced evaluation points $t_k = k/(K-1)$, and compute the total path length as the sum of angular distances between consecutive interpolated rotations:
\begin{equation}
L_{\text{path}} = \sum_{k=1}^{K-1} \|\log(R(t_k)^{-1} R(t_{k+1}))\|_2
\end{equation}
This metric quantifies the total angular displacement along the interpolated path. For ideal geodesic interpolation, $L_{\text{path}}$ equals the shortest path distance on $SO(3)$.

\textbf{Geodesic Deviation:} The relative deviation from the geodesic (shortest path on $SO(3)$) quantifies interpolation optimality:
\begin{equation}
\varepsilon_{\text{geo}} = \frac{|L_{\text{path}} - L_{\text{geodesic}}|}{L_{\text{geodesic}} + \delta}
\end{equation}
where $L_{\text{geodesic}} = \|\log(R_1^{-1} R_2)\|_2$ is the angular distance between the two rotations, and ($\delta > 0$) is a small regularization parameter introduced to avoid division by zero (typically $\delta = 10^{-8}$). Values of $\varepsilon_{\text{geo}}$ close to zero indicate that the interpolation closely follows the geodesic on $SO(3)$.

\textbf{SLERP Implementation:} For quaternion interpolation, we directly implement Eq.~\eqref{eq:SLERP} with the antipodal correction and numerical stability fallback to NLERP when $\upsilon < 0.001$ rad.

\textbf{Derivative Continuity Assessment:} To evaluate the smoothness of the interpolation, we compute the normalized standard deviation of the numerical angular velocities at $K=50$ uniformly distributed interior time points $t_k \in [0.01, 0.99]$ (excluding endpoints):
\begin{equation}
\sigma_{\text{deriv}} = \frac{\text{std}(\{\dot{\omega}_k\}_{k=1}^K)}{\text{mean}(\{\dot{\omega}_k\}_{k=1}^K) + \delta }
\end{equation}
where each angular velocity estimate is computed using central differences:
\begin{equation}
\dot{\omega}_k \approx \frac{\|\log(R(t_k - \frac{\Delta t}{2})^{-1} R(t_k + \frac{\Delta t}{2}))\|_2}{\Delta t} 
\end{equation}
with a fixed time step $\Delta t = 0.001$. Larger $\sigma_{\text{deriv}}$ values indicate variations in angular velocity, potentially leading to jerky motion in animations or instabilities in robotic control.

\textbf{Robustness Analysis:} We assess robustness by evaluating the failure rate and error statistics across a set of $N_{\text{robust}} = 200$ edge cases. These cases include identity rotations, small angles, large angles, gimbal lock configurations, antipodal quaternions, and random rotations. The failure rate is:
\begin{equation}
F_{\text{rate}} = \frac{1}{N_{\text{robust}}} \sum_{i=1}^{N_{\text{robust}}} \mathbf{I} [\varepsilon_i > 0.1 \text{ rad or exception}]
\end{equation}
where $\varepsilon_i$ is the reconstruction error for test case $i$. We additionally compute the mean and maximum angular reconstruction errors across successful conversions:
\begin{equation}
\varepsilon_{\text{avg}} = \frac{1}{|\mathcal{S}|} \sum_{i \in \mathcal{S}} \varepsilon_i, \quad \varepsilon_{\text{max}} = \max_{i \in \mathcal{S}} \varepsilon_i,
\end{equation}
where $\mathcal{S}$ is the set of successful conversions (no exceptions) and $\varepsilon_i = \|\log(\hat{R}_i^{-1} R_i)\|_2$.

\textbf{Computational Efficiency Metrics}

\textbf{Composition Time:} Average time for computing $R_3 = R_1 \circ R_2$ (composition of two rotations) over $N_{\text{time}} = 1000$ trials after 100 warmup iterations:
\begin{equation}
T_{\text{comp}} = \frac{1}{N_{\text{time}}} \sum_{j=1}^{N_{\text{time}}} t_j^{\text{compose}}
\end{equation}
measured in microseconds. For fair comparison, all operations are performed in the native representation (quaternion multiplication for quaternions, matrix multiplication for matrices, etc.).

\textbf{Batch Processing Efficiency:} For vectorization assessment with batch size $B = 100$ rotations processed simultaneously:
\begin{equation}
T_{\text{batch}} = \frac{T_{\text{convert}}^{\text{batch}} + T_{\text{compose}}^{\text{batch}}}{B}
\end{equation}
where $T_{\text{convert}}^{\text{batch}}$ is the time to convert $B$ rotation matrices to the representation and $T_{\text{compose}}^{\text{batch}}$ is the time to compose $B$ pairs of rotations. Lower per-operation times in batch mode indicate better vectorization efficiency.

\begin{table*}[t]
\centering
\caption{Empirical Evaluation Results on Different Metrics}
\label{tab:empirical}
\resizebox{\textwidth}{!}{%
\begin{tabular}{@{}lccccccc@{}}
\toprule
Representation & Storage (bytes) $\downarrow$ & Comp. Time ($\mu$s) $\downarrow$ & Interp. Time ($\mu$s) $\downarrow$ & Path Length $\downarrow$ & Deriv. Continuity $\downarrow$ & Batch Efficiency ($\mu$s) $\downarrow$ \\
\midrule
Euler Angles\cite{euler1776formulae}        & 24 (3 params) & 55.36     & 64.74  & 1.6494 & 0.0014 & 37.68  \\
Axis-Angle\cite{euler1776nova}              & 24 (3 params) & 35.78     & 24.55  & 1.6494 & 0.0014 & 13.27  \\
Quaternions\cite{hamilton1840new}           & 32 (4 params) & 34.25     & 41.18  & 1.6447 & 0.0000 & 14.43  \\
Rotation Matrices\cite{rodrigues1840lois}   & 72 (9 params) & 306.07    & 343.06 & 1.6447 & 0.0000 & 157.02 \\
Exponential Maps\cite{lie1893theorie}       & 24 (3 params) & 19.43     & 24.43  & 1.6494 & 0.0029 & 13.55  \\
6D Continuous\cite{zhou2019continuity}      & 48 (6 params) & 421.95    & 454.62 & 3.7310 & 1.2685 & 227.38 \\
Matrix Fisher\cite{mardia1984maximum}       & 72 (9 params) & N/A       & N/A    & N/A    & N/A    & N/A    \\
\bottomrule
\end{tabular}%
}
\end{table*}

\textbf{Heuristic Evaluations:} We evaluate the following metrics heuristically to assess the efficiency and suitability of various rotation representations, considering memory alignment, hardware optimization, and machine learning compatibility \cite{diebel2006representing, fog2006optimizing, kurzak2009optimizing, murray2017mathematical, martinek2009optimal, williams2019efficacy, grassia1998practical, van2005quaternion, yi2024study, geist2024learning, hempel20226d}:

\begin{itemize}
  \item \textbf{Memory Alignment Score ($A_{\text{mem}}$)} \cite{diebel2006representing, fog2006optimizing, kurzak2009optimizing, murray2017mathematical, martinek2009optimal, williams2019efficacy}: Quantifies memory usage efficiency for SIMD-friendly data structures. Modern CPUs achieve optimal performance with data aligned to cache line boundaries (typically 64 bytes) and SIMD register widths (128, 256, or 512 bits). The score is defined as:
    \begin{equation}
    \resizebox{.89\linewidth}{!}{$
    A_{\text{mem}} = \begin{cases}
    1.0 & \text{if } S_{\text{bytes}} = 32 \text{ (quaternion, 4×64-bit)} \\
    0.9 & \text{if } S_{\text{bytes}} = 24 \text{ (3-param, 3×64-bit)} \\
    0.7 & \text{if } S_{\text{bytes}} = 48 \text{ (6D continuous)} \\
    0.3 & \text{if } S_{\text{bytes}} = 72 \text{ (9-param, poor alignment)} \\
    0.5 & \text{otherwise}
    \end{cases}
    $}
    \end{equation}

  \item \textbf{Hardware Optimization Score ($H_{\text{opt}}$)} \cite{grassia1998practical, van2005quaternion, diebel2006representing, kurzak2009optimizing, murray2017mathematical, williams2019efficacy, yi2024study}: Assesses suitability for vectorized hardware (e.g., SIMD, GPU tensor cores). Assigned values based on operation characteristics: Euler angles (0.6, simple arithmetic but expensive trigonometry), axis-angle (0.8, efficient balance of operations), quaternions (0.9, highly SIMD-friendly 4-component structure), rotation matrices (0.7, well-optimized matrix operations available), exponential map (0.6, requires expensive exp/log operations), 6D continuous (0.5, Gram-Schmidt less amenable to vectorization).
    
  \item \textbf{Machine Learning Compatibility Score ($C_{\text{ML}}$)} \cite{kurzak2009optimizing, murray2017mathematical, williams2019efficacy, geist2024learning, yi2024study, hempel20226d}: Evaluates continuity, differentiability, and parameterization efficiency for gradient-based learning:
    \begin{equation}
    C_{\text{ML}} = w_1 C_{\text{cont}} + w_2 C_{\text{diff}} + w_3 C_{\text{param}}
    \end{equation}
    where:
    \begin{itemize}
    \item $C_{\text{cont}} \in [0,1]$ measures continuity (0 for Euler due to gimbal lock discontinuities, 0.7 for quaternions due to antipodal ambiguity, 1 for truly continuous representations)
    \item $C_{\text{diff}} \in [0,1]$ assesses differentiability and gradient flow quality
    \item $C_{\text{param}} \in [0,1]$ evaluates parameterization efficiency (penalizes overparameterization and constraints)
    \item Equal weights: $w_1 = w_2 = w_3 = 1/3$
    \end{itemize}    
  Assigned composite values: Euler (0.3), axis-angle (0.7), quaternion (0.8, despite double-cover), rotation matrix (0.6, overparameterized), exponential map (0.7), 6D continuous (0.9, specifically designed for deep learning).
\end{itemize}

Together, these metrics form a comprehensive framework for comparing rotation representations in terms of memory efficiency, hardware vectorization compatibility, and suitability for machine-learning applications. The heuristic nature of these scores reflects the fact that real-world performance is influenced by hardware architecture, compiler optimizations, and application-specific usage patterns.

\Cref{tab:empirical} summarizes the key empirical results. Quaternions exhibit low composition times ($\sim 34.25 \, \mu$s) and excellent interpolation properties (path length 1.6447, near-optimal geodesic), with perfect derivative continuity ($\sigma_{\text{deriv}}=0$) using SLERP. Euler angles show longer composition times ($\sim 55.36 \, \mu$s) due to trigonometric evaluations and poor interpolation near singularities. Rotation matrices, while singularity-free, require substantially more computation time (306.07 $\mu$s for composition, 343.06 $\mu$s for interpolation) and storage (72 bytes).

The 6D continuous representation exhibits higher computational costs (421.95 $\mu$s for composition, and 454.62 $\mu$s for interpolation) and produces non-geodesic paths (with a length of 3.7310, significantly greater than the geodesic distance), but achieves the highest ML compatibility score (0.9) due to its continuity properties that facilitate gradient-based optimization. The increased path length arises from the Gram-Schmidt orthogonalization process, which does not preserve geodesic distances when performing linear interpolation in the 6D representation space.

Exponential maps demonstrate excellent computational efficiency (19.43 $\mu$s composition), rivaling quaternions, but show slightly worse derivative continuity (0.0029) due to numerical sensitivities in the exponential/logarithm computations. Axis-angle representations offer a good balance with 35.78 $\mu$s composition time and compact storage (24 bytes).

The matrix Fisher distribution is assessed qualitatively: it provides perfect theoretical continuity and a natural representation of uncertainty. However, it requires expensive sampling operations (typically via rejection sampling or Markov Chain Monte Carlo) and likelihood evaluations involving special functions, which makes quantitative timing comparisons impractical within this framework.

Batch processing results show that quaternions maintain their efficiency advantage even in vectorized operations (14.43 $\mu$s per operation in batches of 100), while rotation matrices suffer from memory bandwidth limitations (157.02 $\mu$s per operation). The 6D representation shows the highest batch processing cost (227.38 $\mu$s) due to the sequential nature of Gram-Schmidt orthogonalization.

\begin{figure}[t]
\centering
\includegraphics[width=0.47\textwidth]{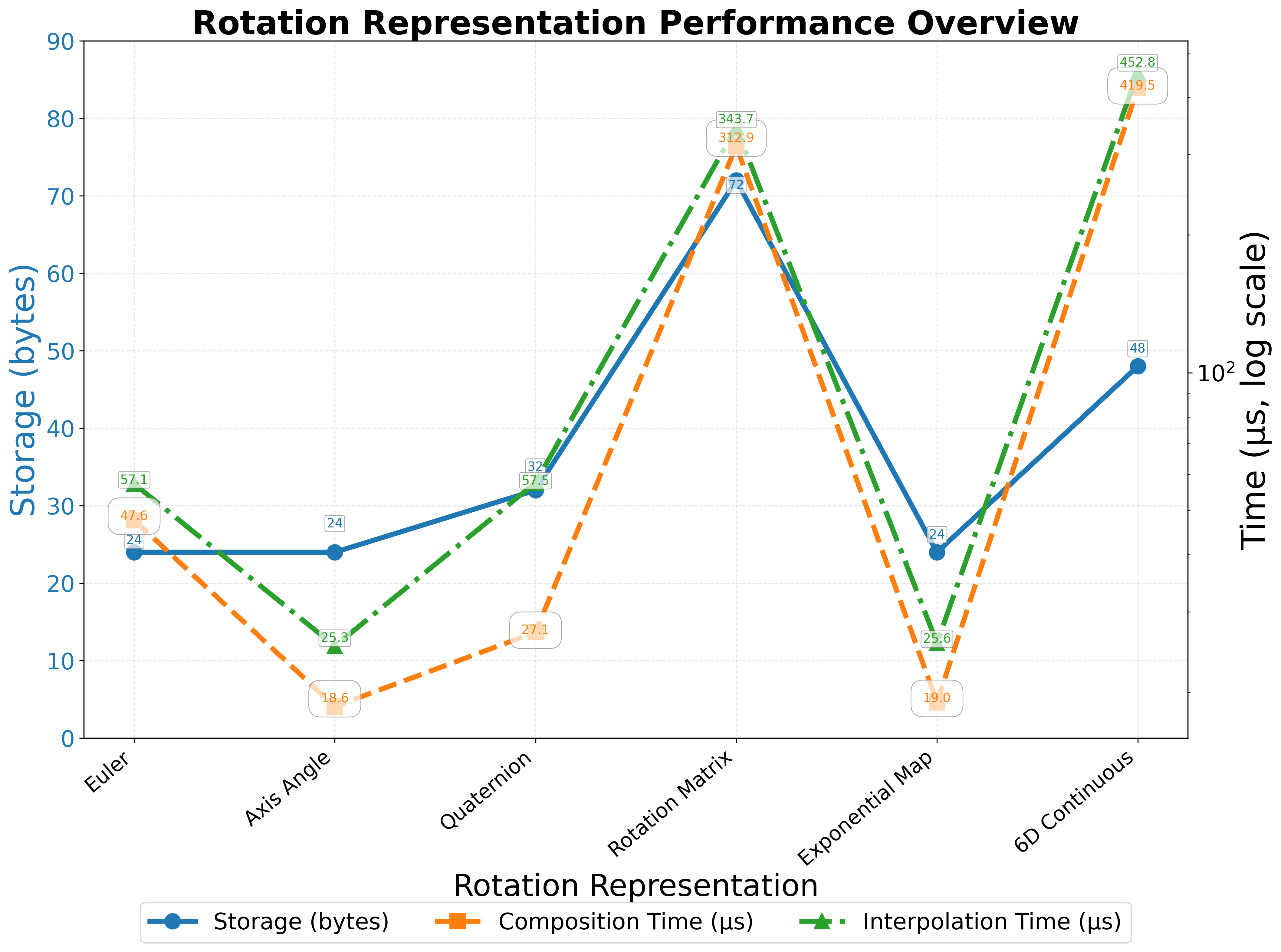}
\caption{Comparison of storage requirements, composition time, and interpolation time for different rotation representations. Quaternions offer an optimal balance across all three metrics.}
\label{fig:storage_performance}
\end{figure}

These results confirm the efficiency of quaternions for general-purpose rotation representation and composition, while 6D representations excel in machine learning contexts where continuity and differentiability are paramount. Visualizations, including the performance-storage-interpolation trade-off in \cref{fig:storage_performance}, further illustrate this multi-dimensional optimization space, where quaternions consistently rank among the top choices across multiple criteria.

\subsection{Pros and Cons Summary}

\Cref{tab:empirical,tab:combined} provide a detailed comparison across multiple dimensions. Quaternions stand out for their computational efficiency, compact storage, and effective interpolation capabilities (via SLERP), but face challenges with continuity in machine learning contexts due to the antipodal ambiguity where $q$ and $-q$ represent the same rotation \cite{zhou2019continuity}. This double-cover property may lead to optimization difficulties in neural networks, as gradient descent can oscillate between equivalent yet parameterically distant representations.

Continuous representations, particularly the 6D formulation, address these continuity shortcomings at the cost of increased storage (48 bytes vs. 32 bytes for quaternions) and computational overhead ($\sim$12× slower composition). Nonetheless, in gradient-based optimization tasks, such as neural network pose estimation, the improved convergence and stability often justify the additional computational cost, with reported error reductions of 6–14\% in autoencoder-based models \cite{zhou2019continuity}.

Probabilistic approaches (matrix Fisher and Bingham distributions) enhance uncertainty modeling, which is critical for safety-critical systems such as autonomous vehicles and surgical robotics, albeit with significantly added computational complexity for sampling and likelihood evaluation \cite{mohlin2020probabilistic}. These methods are most appropriate when explicit uncertainty quantification is required, such as in sensor fusion, where measurement noise must be properly propagated through rotation compositions.

Euler angles remain relevant for human interpretable interfaces and certain aerospace applications where the decomposition into sequential axis rotations has physical meaning (e.g., roll-pitch-yaw). However, their susceptibility to gimbal lock and poor interpolation properties make them unsuitable for general-purpose computational use.

Rotation matrices, despite their overparameterization and computational cost, serve as a universal intermediate representation for conversions and provide a singularity-free formulation that is sometimes preferred in theoretical analysis and certain numerical algorithms where maintaining exact orthogonality through specialized methods (e.g., orthogonal integrators) is feasible.

\begin{table*}[t]
\centering
\caption{Comprehensive Comparison of Rotation Representations}
\label{tab:combined}
\resizebox{\textwidth}{!}{%
\begin{tabular}{@{}lccccc@{}}
\toprule
Representation & Gimbal Lock & Storage (bytes) & Composition Operations & Interpolation Method & Continuity \\
\midrule
Euler Angles\cite{euler1776formulae} & Yes (at \(\pm \pi/2\)) & 24 & Trig. evaluations (6 sin/cos) & Linear (poor near poles) & Discontinuous (gimbal lock) \\
Axis-Angle\cite{euler1776nova} & Yes (at \(2\pi k\)) & 24 & 27 mult/18 add (via matrix) & Linear (non-geodesic) & Continuous (except at $\theta=k\pi$) \\
Quaternions\cite{hamilton1840new} & No & 32 & 16 mult/12 add (28 FLOPs) & SLERP (geodesic) & Continuous (antipodal ambiguity) \\
Rotation Matrices\cite{rodrigues1840lois} & No & 72 & 27 mult/18 add (45 FLOPs) & Geodesic (via log/exp) & Continuous (with normalization) \\
Exponential Maps\cite{lie1893theorie} & No (local) & 24 & Matrix exp (varies) & Linear (non-geodesic) & Continuous (singularities at $\theta=k\pi$) \\
6D Continuous\cite{zhou2019continuity} & No & 48 & Gram-Schmidt + mult & Linear then orthogonalize & Fully continuous \\
Matrix Fisher\cite{mardia1984maximum} & No & 72 & Probabilistic sampling & Bayesian blending & Continuous (distributional) \\
\bottomrule
\end{tabular}%
}
\end{table*}

\section{Applications}
\label{sec:applications}

The choice of rotation representation significantly impacts performance and reliability in practical applications. We survey key application domains and provide evidence-based recommendations.

\subsection{Computer Graphics and Animation}
Shoemake \cite{shoemake1985animating} pioneered the use of quaternions in keyframe animation, demonstrating that SLERP effectively prevents gimbal lock during complex camera maneuvers (e.g., barrel rolls in flight simulators) and enables smooth rotational transitions. The constant angular velocity property of SLERP produces visually natural motion, which is crucial for character animation and camera path planning.

Hart et al. \cite{hart1994visualizing} developed interactive visualizations of quaternion "belts" (the phenomenon where a 720-degree rotation, not 360 degrees, returns a quaternion to its original value), aiding educational understanding of the topology of rotation spaces and the double-cover relationship between quaternions and $SO(3)$. These visualizations have become standard in computer graphics curricula.

Exponential maps prove advantageous in physics-based simulations, enabling smooth integration of rotational dynamics through their natural connection to angular velocities. In rigid body dynamics, the time derivative of a rotation can be expressed compactly in the exponential map parameterization as 
\begin{equation}
\dot{\mathbf{v}} = J(\mathbf{v})\boldsymbol{\omega},
\end{equation} 
where $\boldsymbol{\omega}$ is the angular velocity and $J(\mathbf{v})$ is a Jacobian matrix, facilitating accurate motion tracking and stable numerical integration \cite{grassia1998practical}.

Modern game engines and animation software (e.g., Unity, Unreal Engine, Blender) predominantly use quaternions for internal rotation representation due to their computational efficiency and robust interpolation properties, while exposing Euler angles in user interfaces for intuitive manual control. This hybrid approach leverages the strengths of both representations.

\subsection{Robotics and Navigation}
In inertial navigation systems (INS), exponential maps efficiently handle incremental rotational updates from gyroscope measurements, making them ideal for real-time applications \cite{brossard2019rins}. The small-angle approximation 
\begin{equation}
R \approx I + [\Delta\mathbf{v}]_\times
\end{equation} 
enables fast integration of angular velocity measurements:
\begin{equation}
\mathbf{v}_{t+\Delta t} = \mathbf{v}_t + \boldsymbol{\omega}_t \Delta t,
\end{equation} 
where $\boldsymbol{\omega}_t$ is the measured angular velocity. This approach maintains computational efficiency while achieving adequate accuracy for typical sensor update rates (100–1000 Hz).

Quaternions are widely adopted in Simultaneous Localization and Mapping (SLAM) systems due to their compact storage (critical for storing large pose graphs), rapid composition for pose chain multiplication, and numerical stability. Modern visual-inertial SLAM systems (e.g., ORB-SLAM, VINS-Mono) use quaternions for camera pose representation, achieving real-time performance on embedded platforms. The Extended Kalman Filter (EKF) and Unscented Kalman Filter (UKF) formulations for orientation estimation commonly use quaternion parameterizations to avoid gimbal lock during aggressive maneuvers.

Matrix Fisher distributions enhance uncertainty modeling in sensor fusion processes, particularly in multi-sensor systems combining IMU, GPS, and visual odometry. By representing orientation estimates as probability distributions rather than point estimates, these methods properly account for measurement noise, temporal correlation, and sensor-specific uncertainty characteristics, improving reliability in challenging environments such as GPS-denied indoor spaces or visually low-texture scenes \cite{mohlin2020probabilistic}. The concentration parameters of the Fisher distribution can be adapted online based on sensor quality metrics, enabling robust fusion even with intermittent sensor failures.

In robotic manipulation, quaternions help smooth trajectory planning and control. The SLERP interpolation ensures constant angular velocity in joint space or task space orientation planning, preventing jerky motions that could excite structural resonances or violate actuator limits. For applications requiring higher-order continuity (e.g., $C^2$ continuous acceleration profiles), quaternion splines (e.g., squad curves, B-splines on $\mathbb{S}^3$) extend SLERP to provide smooth acceleration profiles essential for high-speed pick-and-place operations.

\subsection{Machine Learning and Vision}
Zhou et al. \cite{zhou2019continuity} demonstrated that 6D continuous representations reduce pose estimation errors by 6–14\% compared to quaternions in autoencoder-based models and convolutional neural networks for single-image pose estimation. This improvement is attributed to the continuous, bijective mapping to $SO(3)$, which eliminates discontinuities in the loss landscape that can trap gradient-based optimizers in local minima or cause training instabilities.

In practice, the 6D representation enables neural networks to learn rotation-related tasks more effectively by providing smooth gradients throughout the entire rotation space. The Gram-Schmidt orthogonalization in Equations~\eqref{eq:6DRep1}-\eqref{eq:6DRep3} is differentiable almost everywhere (excluding the measure-zero set where inputs are linearly dependent), allowing backpropagation through the rotation recovery process. Empirical studies show faster convergence (20–30\% fewer training epochs) and improved generalization on validation sets compared to quaternion-based networks, which must address the antipodal ambiguity.

Probabilistic approaches reinforce robustness against noisy data in head pose estimation, where Bingham distributions model uncertainty in facial landmark detection and the resulting orientation estimates \cite{gilitschenski2015unscented}. By propagating uncertainty through the estimation pipeline, these methods provide confidence bounds on predictions, enabling downstream systems to appropriately weight pose estimates based on image quality, occlusions, and other reliability indicators.

Recent advances in equivariant neural networks for 3D data (point clouds, meshes) leverage rotation-equivariant architectures that process rotations in their native $SO(3)$ representation or through spherical harmonics. These networks often utilize rotation matrices or quaternions internally, with carefully designed architectures that ensure the outputs transform consistently under input rotations. Such properties are critical for applications including 3D object recognition, scene understanding, and molecular property prediction in computational chemistry.

In multi-view geometry and structure from motion (SfM), rotation averaging problems (finding the best set of absolute rotations given noisy relative rotations between views) benefit from exponential map parameterizations. The geodesic distance on $SO(3)$ naturally corresponds to the rotation angle, making it an appropriate metric for least-squares optimization:
\begin{equation}
\min_{\{R_i\}} \sum_{(i,j) \in \mathcal{E}} \|\log(R_{ij}^{-1} R_i^{-1} R_j)\|_2^2,
\end{equation}
where $R_{ij}$ are measured relative rotations and $\mathcal{E}$ is the set of image pairs.

\subsection{3D Shape Registration}
Quaternions ease efficient least-squares registration of two corresponding point sets \(\{p_i\}_{i=1}^N\) and \(\{q_i\}_{i=1}^N\) by optimizing the alignment transformation. The objective function to minimize is:
\begin{equation}
E(R, \mathbf{t}) = \sum_{i=1}^N \|R p_i + \mathbf{t} - q_i\|^2,    
\end{equation}
where $R \in SO(3)$ is the rotation and $\mathbf{t} \in \mathbb{R}^3$ is the translation. This problem can be decoupled by first removing the centroids. 

The registration process involves:
\begin{enumerate}
\item Computing centroids: $\bar{p} = \frac{1}{N}\sum_{i=1}^N p_i$, $\bar{q} = \frac{1}{N}\sum_{i=1}^N q_i$ and defining center points: $p_i' = p_i - \bar{p}$ and $q_i' = q_i - \bar{q}$.
\item Constructing the cross-covariance matrix: $H = \sum_{i=1}^N p_i' (q_i')^\top$.
\item Building the symmetric matrix $M$ from $H$:
\begin{equation}
\scalebox{0.85}{$
M = \begin{pmatrix}
\text{tr}(H) & \Delta^\top \\
\Delta & H + H^\top - \text{tr}(H)I
\end{pmatrix}
$}
\end{equation}
where $\Delta = (H_{23} - H_{32}, H_{31} - H_{13}, H_{12} - H_{21})^\top$.
\item Finding the optimal rotation: the eigenvector $\mathbf{q}^* = (q_0, q_1, q_2, q_3)^\top$ corresponding to the largest eigenvalue of $M$ gives the optimal unit quaternion
\item Computing the optimal translation: $\mathbf{t}^* = \bar{q} - R^* \bar{p}$, where $R^*$ is the rotation matrix obtained from $\mathbf{q}^*$
\end{enumerate}

This quaternion-based formulation avoids iterative optimization over rotation parameters, yielding a closed-form solution via eigenvalue decomposition. The method is known as Horn's algorithm \cite{horn1987closed} and achieves global optimality for the given point correspondences.

The Iterative Closest Point (ICP) algorithm \cite{besl1992method} extends this method to handle cases with unknown point correspondences. ICP alternates between:
\begin{enumerate}
\item Finding nearest-neighbor correspondences between source and target point clouds
\item Solving the registration problem using the quaternion method above
\item Transforming the source points and iterating until convergence
\end{enumerate}

Modern variants include point-to-plane ICP (which uses surface normals to improve convergence) and probabilistic ICP (which models correspondence uncertainty). Quaternions remain the preferred rotation parameterization due to their efficiency in the inner loop, where millions of registrations may be performed across multi-scale pyramid levels.

In deformable shape registration and non-rigid structure-from-motion, per-point or per-region rotations must be optimized jointly with spatial regularization terms. Quaternion parameterizations enable efficient gradient-based optimization with smooth energy landscapes, while continuous 6D representations are promising for learning-based deformable registration, where neural networks predict dense rotation fields.

\subsection{Other Applications}

\subsubsection{Physics and Relativity}
Quaternions play an instrumental role in modeling Lorentz transformations in special relativity, where they provide a compact representation of boosts and rotations in spacetime. The quaternion formulation of relativistic mechanics enables efficient computation of particle trajectories and enables geometric interpretations of relativistic phenomena. Spinors, closely related to quaternions, are fundamental in quantum mechanics for representing particle spin states.

\subsubsection{Signal Processing}
The Quaternion Fourier Transform (QFT) extends classical Fourier analysis to quaternion-valued signals, supporting advanced image processing techniques. Applications include:
\begin{itemize}
\item Color image watermarking: QFT processes color channels holistically rather than independently, preserving inter-channel correlations and improving robustness against attacks
\item Color image recognition: Quaternion neural networks leverage the QFT for feature extraction in color texture analysis and object recognition
\item Multidimensional signal analysis: Vector-valued signals (e.g., RGB images, wind velocity fields, electromagnetic fields) can be processed using quaternion algebra, enabling phase-aware filtering and analysis
\end{itemize}

The hypercomplex structure of quaternions enables simultaneous processing of multiple signal components while preserving inter-component relationships, offering advantages over component-wise approaches in specific applications.

\subsubsection{Aerospace Engineering}
Euler angles maintain significance in aerospace applications for their intuitive correspondence to physical rotation sequences. Aircraft attitude is commonly described using roll-pitch-yaw (equivalent to an $XYZ$ Euler sequence), where each angle has direct physical meaning:
\begin{itemize}
\item Roll: rotation about the longitudinal axis (forward direction)
\item Pitch: rotation about the lateral axis (wing-to-wing direction)
\item Yaw: rotation about the vertical axis
\end{itemize}

However, flight control systems usually use quaternions internally for computation, converting to Euler angles only for display purposes or when interfacing with human operators. This hybrid approach avoids gimbal lock in the control loops while maintaining interpretable representations for pilots and engineers.

\subsubsection{Medical Imaging and Surgical Robotics}
In medical image registration (e.g., aligning CT and MRI scans), rotation representations must balance computational efficiency with robustness. Quaternions are commonly used for rigid registration, while probabilistic methods (Fisher or Bingham distributions) provide uncertainty estimates critical for surgical planning. Robot-assisted surgery systems employ quaternions to achieve smooth and predictable instrument motion, where SLERP ensures near-constant angular velocity during automated tool positioning.

\begin{figure}[t]
\centering
\includegraphics[width=0.47\textwidth]{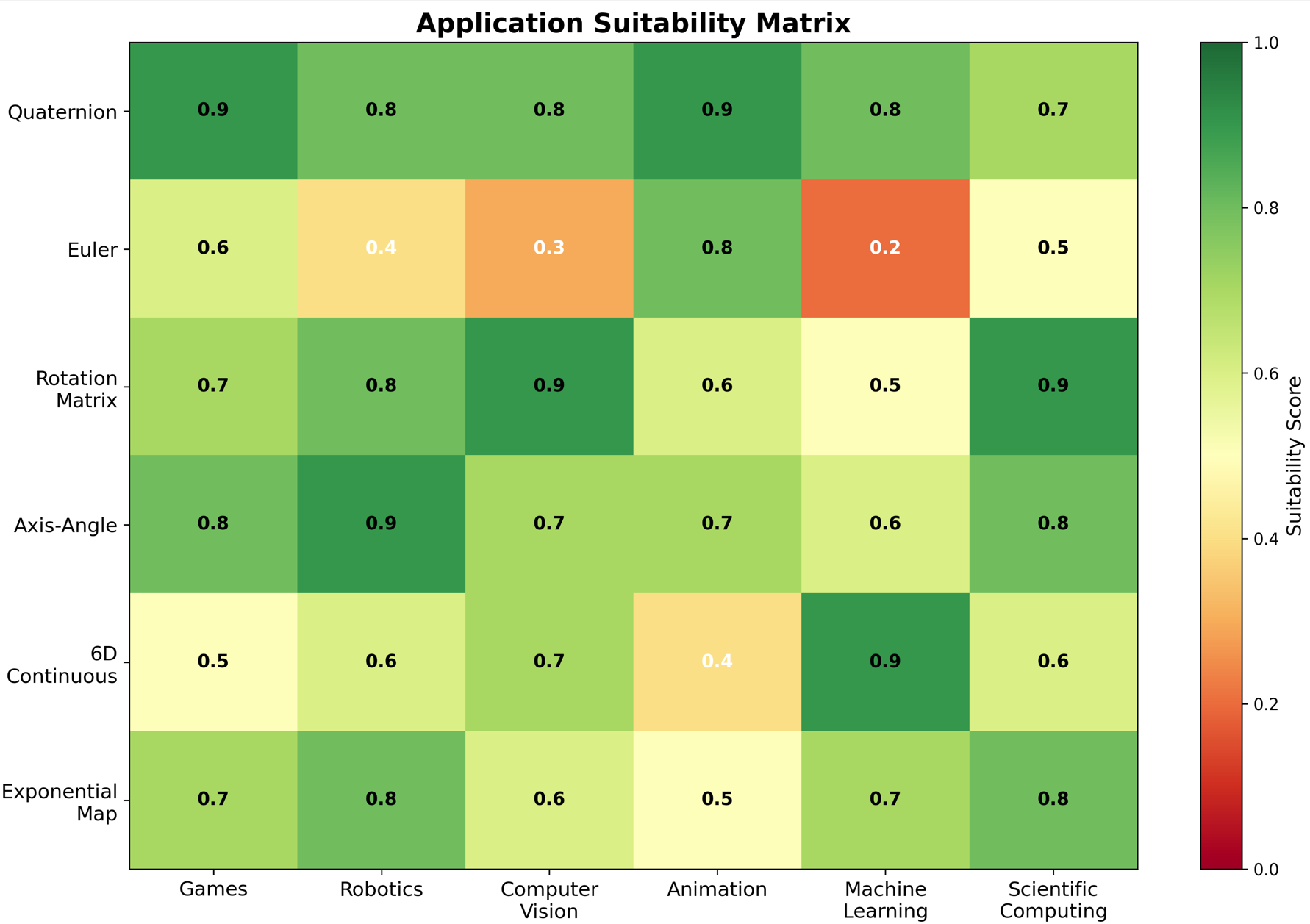}
\caption{Correlation map showing the applicability of different rotation representations across application domains. Darker shading indicates stronger suitability based on empirical metrics and application-specific requirements. Quaternions demonstrate broad applicability, while specialized representations excel in targeted domains.}
\label{fig:application_mat}
\end{figure}

\Cref{fig:application_mat} provides a correlation map showing the suitability of different rotation representations across application domains based on our empirical evaluation metrics and literature analysis. The heatmap indicates that quaternions show strong applicability across most domains (darker shading), while specialized representations like 6D continuous excel specifically in machine learning contexts, and probabilistic methods are most valuable in uncertainty-critical applications such as sensor fusion and safety-critical robotics.

\section{Future Directions}
\label{sec:future}

Despite extensive research on rotation representations, several promising directions remain underexplored and warrant further investigation.

\subsection{Hybrid Representations}
Future research should systematically explore hybrid representations that combine the strengths of different approaches. For instance, combining quaternions with 6D continuous methods could leverage quaternions' computational efficiency for composition and interpolation while using 6D representations at neural network boundaries (input/output layers) to facilitate learning. Such hybrid architectures could potentially achieve both fast inference and effective training. Recent work by Pepe et al. \cite{pepe2024learning} has demonstrated that parametrizing rotations using geometric algebra (GA) as bivectors can outperform the 6D representation while requiring fewer parameters and offering enhanced robustness to noise, suggesting that GA provides a broader framework for describing rotations that is particularly suitable for regression tasks via deep learning.

Adaptive representation selection based on runtime conditions (e.g., switching between exponential maps for small rotations and quaternions for large rotations) could optimize performance across diverse operating regimes. Dynamic selection strategies using machine learning to predict the most efficient representation for a given task instance represent another promising direction.

\subsection{Higher-Dimensional Generalizations}
Exploring n-dimensional generalizations for $SO(n)$ in graphics and high-dimensional data analysis could yield valuable insights. While quaternions elegantly handle $SO(3)$, their generalization to higher dimensions (e.g., octonions for limited aspects of $SO(7)$) remains limited in practical utility. Alternative approaches such as Clifford algebras provide a unified framework for rotations in arbitrary dimensions, but their computational properties and practical applicability require further study.

Comparing these algebraic approaches with homogeneous coordinates in projective geometry could elucidate their relative merits. Note that homogeneous coordinates cannot directly incorporate quaternion-based transformations for rotations with translations due to their distinct algebraic properties: quaternions operate in a four-dimensional division algebra while homogeneous coordinates embed operations in projective space. Developing unified frameworks that combine the benefits of both approaches (e.g., dual quaternions for rigid body transformations) deserves attention. Recent advances in dual quaternion operations for rigid body motion have been successfully applied to robotic kinematics, with Wang et al. \cite{wang2024dual} proposing novel simultaneous methods for hand-eye calibration that verify how rotation problems can be extended to rigid-body transformation problems via dual algebra.

\subsection{Standardized Benchmarking}
The field would benefit significantly from standardized benchmark suites for evaluating rotation representations across diverse applications. Such benchmarks should include:
\begin{itemize}
\item Canonical test problems with ground truth solutions for validation
\item Diverse operating conditions (small/large angles, near-singularities, random rotations)
\item Multiple performance metrics (accuracy, speed, memory, numerical stability)
\item Various computational platforms (CPU, GPU, embedded systems)
\item Application-specific scenarios (animation, SLAM, pose estimation, registration)
\end{itemize}

Public repositories with reference implementations in multiple programming languages would help reproducible research and enable practitioners to make informed choices based on empirical evidence rather than conventional wisdom. The computer vision community's benchmark datasets (e.g., ModelNet for 3D shape analysis, KITTI for SLAM) could be extended with rotation-specific evaluation protocols. Recent research has demonstrated the practical benefits of proper rotation representation selection in deep learning applications, with Hempel et al. \cite{hempel20226d} showing that continuous 6D rotation matrix representations enable robust direct regression for head pose estimation, outperforming state-of-the-art methods by nearly 20\% on the AFLW2000 benchmark while avoiding the discontinuities and ambiguities inherent in Euler angle and quaternion representations.

\subsection{Learning-Based Representation Selection}
Machine learning methods could be developed to automatically select optimal rotation representations based on specific problem characteristics. A meta-learning system could be trained on a diverse corpus of rotation-related tasks to predict which representation will perform best for a new problem instance, considering factors such as:
\begin{itemize}
\item Expected rotation magnitudes and distributions
\item Computational constraints (latency, throughput, memory)
\item Required precision and numerical stability
\item Presence of singularities or problematic regions in the operational space
\item Hardware characteristics (SIMD capabilities, cache hierarchy)
\end{itemize}

Such systems could provide decision support for engineers and researchers, democratizing access to rotation representation expertise.

\subsection{Probabilistic Representation Advances}
Further development of probabilistic rotation representations could enhance uncertainty quantification in safety-critical applications. Recent work by Zhang et al. \cite{zhang2023revisiting} has proposed better modeling of underlying noise distributions by directly propagating uncertainty from point correspondences into rotation averaging, demonstrating improved robustness in rotation estimation tasks. Challenges include:
\begin{itemize}
\item Efficient sampling and inference algorithms for complex distributions over $SO(3)$
\item Tractable composition and marginalization operations on rotation distributions
\item Integration with modern probabilistic programming frameworks
\item Theoretical guarantees on uncertainty calibration and coverage
\end{itemize}

Combining probabilistic representations with deep learning (e.g., probabilistic neural networks that output rotation distributions rather than point estimates) could improve robustness in applications with inherent ambiguity or unreliable inputs.

\subsection{Quantum Computing Applications}
As quantum computing matures, investigating rotation representations suitable for quantum algorithms could yield insights. Quantum state spaces naturally involve complex-valued rotations on high-dimensional spheres, and understanding optimal representations for quantum rotation gates, adiabatic evolution, and quantum simulation of physical systems with rotational symmetry represents an emerging frontier.

\subsection{Geometric Deep Learning}
The intersection of rotation representations with geometric deep learning on manifolds deserves deeper exploration. Developing neural network architectures that natively operate on $SO(3)$ or other rotation groups, without requiring embedding in Euclidean space, could provide benefits for tasks with inherent rotational structure. Recent advances in this area include the work of Gerken et al. \cite{gerken2023geometric}, who surveyed the mathematical foundations of geometric deep learning with focus on group equivariant and gauge equivariant neural networks, demonstrating how group equivariant layers can be interpreted as intertwiners between induced representations and showing their relation to gauge equivariant convolutional layers for spherical networks corresponding to $SO(3)/ SO(2)$. Additionally, Batzner et al. \cite{batzner20223} developed E(3)-equivariant neural network approaches (NequIP) that employ equivariant convolutions for interactions of geometric tensors, achieving state-of-the-art accuracy with remarkable data efficiency—outperforming existing models with up to three orders of magnitude fewer training data. Such architectures might leverage tools from differential geometry, Lie group theory, and harmonic analysis (e.g., $SO(3)$-equivariant convolutions using Wigner D-matrices).

\section{Conclusion}
\label{sec:conclusion}

This paper has provided a comprehensive investigation of rotation representations for the special orthogonal group $SO(3)$, examining their mathematical foundations, computational properties, and practical applications. Through rigorous empirical evaluation and extensive literature synthesis, we have elucidated the fundamental trade-offs inherent in different representation choices.

Quaternions emerge as the dominant general-purpose representation, offering an optimal balance of computational efficiency (34.25 $\mu$s composition time), compact storage (32 bytes), robust interpolation via SLERP (achieving geodesic paths with perfect derivative continuity), and absence of gimbal lock. Their widespread adoption in robotics, computer graphics, and navigation systems is well-justified by these properties. The quaternion-based closed-form solution for 3D point set registration exemplifies their practical utility in geometric algorithms.

However, the quaternion double-cover property (where $q$ and $-q$ represent identical rotations) creates challenges in gradient-based optimization, motivating the development of continuous representations. The 6D continuous representation addresses this limitation, providing a truly continuous, differentiable parameterization that reduces pose estimation errors by 6-14\% in neural network applications. While computationally more expensive (421.95 $\mu$s composition time), this cost is often justified in machine learning contexts where continuity and smooth loss landscapes are paramount.

Probabilistic representations (matrix Fisher and Bingham distributions) extend rotation modeling to inherently uncertain scenarios, enabling principled uncertainty quantification essential for safety-critical systems. Though computationally demanding, these methods provide unique capabilities for sensor fusion, multi-hypothesis tracking, and risk-aware decision-making in autonomous systems.

Traditional representations retain niche relevance: Euler angles for human-interpretable interfaces in aerospace, axis-angle/exponential maps for physics-based simulation and small-angle incremental updates, and rotation matrices as universal intermediates for conversions and theoretical analysis.
Our results, summarized in Tables~\ref{tab:empirical} and \ref{tab:combined}, offer quantitative guidance for representation selection based on application requirements.

Looking forward, hybrid methods that adaptively combine representations, standardized benchmarks enabling reproducible comparison, and integration with emerging computational paradigms (geometric deep learning, quantum computing) represent promising research directions. The choice of rotation representation should be tailored to specific application contexts, with empirical validation serving as the cornerstone for advancing the field.

As computational demands grow in domains ranging from real-time robotics to large-scale 3D reconstruction, continued innovation in rotation representations will remain essential. This paper provides both a comprehensive reference for current best practices and a foundation for future developments in this fundamental area of computational geometry and applied mathematics.

\newpage
{\small
\bibliographystyle{IEEETran}
\bibliography{IEEEfull}

@book{voight2021quaternion,
  title={Quaternion algebras},
  author={Voight, John},
  year={2021},
  publisher={Springer Nature}
}

@inproceedings{shoemake1985animating,
  title={Animating rotation with quaternion curves},
  author={Shoemake, Ken},
  booktitle={Proceedings of the 12th annual conference on Computer graphics and interactive techniques},
  pages={245--254},
  year={1985}
}

@article{hart1994visualizing,
  title={Visualizing quaternion rotation},
  author={Hart, John C and Francis, George K and Kauffman, Louis H},
  journal={ACM Transactions on Graphics (TOG)},
  volume={13},
  number={3},
  pages={256--276},
  year={1994},
  publisher={ACM New York, NY, USA}
}

@article{ten2020let,
  title={Let’s remove quaternions from every 3d engine},
  author={ten Bosch, Marc},
  journal={URL: https://marctenbosch. com/quaternions},
  year={2020}
}

@article{fuhua2023dont,
  title = "We Don’t Really Need Quaternions in Geometric Modeling, Computer Graphics and Animation: Here Is Why",
  author = "Fuhua Cheng and T. Lee Johnson and Anastasia Kazadi and Ethan G. Toney and Jonathan I. Watson and Alice J. Lin",
  note = "Publisher Copyright: © 2023 CAD Solutions, LLC.",
  year = "2023",
  doi = "10.14733/cadaps.2023.1061-1073",
  journal={Computer-Aided Design and Applications},
  volume = "20",
  pages = "1061--1073",
  number = "6"
}

@book{goldman2022rethinking,
  title={Rethinking quaternions},
  author={Goldman, Ron},
  year={2022},
  publisher={Springer Nature}
}

@inproceedings{zhou2019continuity,
  title={On the continuity of rotation representations in neural networks},
  author={Zhou, Yi and Barnes, Connelly and Lu, Jingwan and Yang, Jimei and Li, Hao},
  booktitle={Proceedings of the IEEE/CVF conference on computer vision and pattern recognition},
  pages={5745--5753},
  year={2019}
}

@inproceedings{brossard2019rins,
  title={RINS-W: Robust inertial navigation system on wheels},
  author={Brossard, Martin and Barrau, Axel and Bonnabel, Silvere},
  booktitle={2019 IEEE/RSJ International Conference on Intelligent Robots and Systems (IROS)},
  pages={2068--2075},
  year={2019},
  organization={IEEE}
}

@article{mohlin2020probabilistic,
  title={Probabilistic orientation estimation with matrix fisher distributions},
  author={Mohlin, David and Sullivan, Josephine and Bianchi, G{\'e}rald},
  journal={Advances in Neural Information Processing Systems},
  volume={33},
  pages={4884--4893},
  year={2020}
}

@article{grassia1998practical,
  title={Practical parameterization of rotations using the exponential map},
  author={Grassia, F Sebastian},
  journal={Journal of graphics tools},
  volume={3},
  number={3},
  pages={29--48},
  year={1998},
  publisher={Taylor \& Francis}
}

@article{gilitschenski2015unscented,
  title={Unscented orientation estimation based on the Bingham distribution},
  author={Gilitschenski, Igor and Kurz, Gerhard and Julier, Simon J and Hanebeck, Uwe D},
  journal={IEEE Transactions on Automatic Control},
  volume={61},
  number={1},
  pages={172--177},
  year={2015},
  publisher={IEEE}
}

@inproceedings{besl1992method,
  title={Method for registration of 3-D shapes},
  author={Besl, Paul J and McKay, Neil D},
  booktitle={Sensor fusion IV: control paradigms and data structures},
  volume={1611},
  pages={586--606},
  year={1992},
  organization={Spie}
}

@article{bingham1974antipodally,
  title={An antipodally symmetric distribution on the sphere},
  author={Bingham, Christopher},
  journal={The Annals of Statistics},
  pages={1201--1225},
  year={1974},
  publisher={JSTOR}
}

@inproceedings{mayhew2011quaternion,
  title={On quaternion-based attitude control and the unwinding phenomenon},
  author={Mayhew, Christopher G and Sanfelice, Ricardo G and Teel, Andrew R},
  booktitle={Proceedings of the 2011 american control conference},
  pages={299--304},
  year={2011},
  organization={IEEE}
}

@article{pavllo2018quaternet,
  title={Quaternet: A quaternion-based recurrent model for human motion},
  author={Pavllo, Dario and Grangier, David and Auli, Michael},
  journal={arXiv preprint arXiv:1805.06485},
  year={2018}
}

@article{sommer2018and,
  title={Why and how to avoid the flipped quaternion multiplication},
  author={Sommer, Hannes and Gilitschenski, Igor and Bloesch, Michael and Weiss, Stephan and Siegwart, Roland and Nieto, Juan},
  journal={Aerospace},
  volume={5},
  number={3},
  pages={72},
  year={2018},
  publisher={MDPI}
}

@article{huo2023adaptive,
  title={Adaptive sliding mode attitude tracking control for rigid spacecraft considering the unwinding problem},
  author={Huo, Baoyu and Du, Mingjun and Yan, Zhiguo},
  journal={Mathematics},
  volume={11},
  number={20},
  pages={4372},
  year={2023},
  publisher={MDPI}
}

@article{euler1776formulae,
  title={Formulae generales pro translatione quacunque corporum rigidorum},
  author={Euler, Leonhard},
  journal={Novi commentarii academiae scientiarum petropolitanae},
  pages={189--207},
  year={1776}
}

@article{euler1776nova,
  title={Nova methodus motum corporum rigidorum degerminandi},
  author={Euler, Leonhard},
  journal={Novi commentarii academiae scientiarum Petropolitanae},
  pages={208--238},
  year={1776}
}

@article{hamilton1840new,
  title={On a new species of imaginary quantities, connected with the theory of quaternions},
  author={Hamilton, William Rowan},
  journal={Proceedings of the Royal Irish Academy (1836-1869)},
  volume={2},
  pages={424--434},
  year={1840},
  publisher={JSTOR}
}

@article{rodrigues1840lois,
  title={Des lois g{\'e}om{\'e}triques qui r{\'e}gissent les d{\'e}placements d'un syst{\`e}me solide dans l'espace, et de la variation des coordonn{\'e}es provenant de ces d{\'e}placements consid{\'e}r{\'e}s ind{\'e}pendamment des causes qui peuvent les produire},
  author={Rodrigues, Olinde},
  journal={Journal de math{\'e}matiques pures et appliqu{\'e}es},
  volume={5},
  pages={380--440},
  year={1840}
}

@book{lie1893theorie,
  title={Theorie der transformationsgruppen},
  author={Lie, Sophus and Engel, Friedrich},
  volume={3},
  year={1893},
  publisher={Teubner}
}

@article{mardia1984maximum,
  title={Maximum likelihood estimation of models for residual covariance in spatial regression},
  author={Mardia, Kanti V and Marshall, Roger J},
  journal={Biometrika},
  volume={71},
  number={1},
  pages={135--146},
  year={1984},
  publisher={Oxford University Press}
}

@article{yi2024study,
  title={A Study of Performance Programming of CPU, GPU accelerated Computers and SIMD Architecture},
  author={Yi, Xinyao},
  journal={arXiv preprint arXiv:2409.10661},
  year={2024}
}

@article{diebel2006representing,
  title={Representing attitude: Euler angles, unit quaternions, and rotation vectors},
  author={Diebel, James and others},
  journal={Matrix},
  volume={58},
  number={15-16},
  pages={1--35},
  year={2006}
}

@book{murray2017mathematical,
  title={A mathematical introduction to robotic manipulation},
  author={Murray, Richard M and Li, Zexiang and Sastry, S Shankar},
  year={2017},
  publisher={CRC press}
}

@article{van2005quaternion,
  title={From quaternion to matrix and back},
  author={Van Waveren, JMP},
  journal={Id Software, Inc},
  year={2005}
}

@article{kurzak2009optimizing,
  title={Optimizing matrix multiplication for a short-vector SIMD architecture--CELL processor},
  author={Kurzak, Jakub and Alvaro, Wesley and Dongarra, Jack},
  journal={Parallel Computing},
  volume={35},
  number={3},
  pages={138--150},
  year={2009},
  publisher={Elsevier}
}

@article{williams2019efficacy,
  title={On the efficacy and high-performance implementation of quaternion matrix multiplication},
  author={Williams-Young, David and Li, Xiaosong},
  journal={arXiv preprint arXiv:1903.05575},
  year={2019}
}

@inproceedings{geist2024learning,
  title={Learning with 3D rotations: a Hitchhiker's guide to SO (3)},
  author={Geist, A Ren{\'e} and Frey, Jonas and Zhobro, Mikel and Levina, Anna and Martius, Georg},
  booktitle={Proceedings of the 41st International Conference on Machine Learning},
  pages={15331--15350},
  year={2024}
}

@article{fog2006optimizing,
  title={Optimizing software in C++},
  author={Fog, Agner},
  journal={URL: http://www. agner. org/optimize/optimizing\_cpp. pdf},
  year={2006}
}

@article{martinek2009optimal,
  title={Optimal rotation alignment of 3D objects using a GPU-based similarity function},
  author={Martinek, Michael and Grosso, Roberto},
  journal={Computers \& Graphics},
  volume={33},
  number={3},
  pages={291--298},
  year={2009},
  publisher={Elsevier}
}

@inproceedings{cao2021vector,
  title={A vector-based representation to enhance head pose estimation},
  author={Cao, Zhiwen and Chu, Zongcheng and Liu, Dongfang and Chen, Yingjie},
  booktitle={Proceedings of the IEEE/CVF Winter Conference on applications of computer vision},
  pages={1188--1197},
  year={2021}
}

@inproceedings{hempel20226d,
  title={6d rotation representation for unconstrained head pose estimation},
  author={Hempel, Thorsten and Abdelrahman, Ahmed A and Al-Hamadi, Ayoub},
  booktitle={2022 IEEE International Conference on Image Processing (ICIP)},
  pages={2496--2500},
  year={2022},
  organization={IEEE}
}

@article{kim2023rotation,
  title={Rotation representations and their conversions},
  author={Kim, Soohwan and Kim, Minkyoung},
  journal={Ieee Access},
  volume={11},
  pages={6682--6699},
  year={2023},
  publisher={IEEE}
}

@article{horn1987closed,
  title={Closed-form solution of absolute orientation using unit quaternions},
  author={Horn, Berthold KP},
  journal={Journal of the optical society of America A},
  volume={4},
  number={4},
  pages={629--642},
  year={1987},
  publisher={Optical Society of America}
}

@article{sei2013properties,
  title={Properties and applications of Fisher distribution on the rotation group},
  author={Sei, Tomonari and Shibata, Hiroki and Takemura, Akimichi and Ohara, Katsuyoshi and Takayama, Nobuki},
  journal={Journal of Multivariate Analysis},
  volume={116},
  pages={440--455},
  year={2013},
  publisher={Elsevier}
}

@article{lee2018bayesian,
  title={Bayesian attitude estimation with the matrix Fisher distribution on SO (3)},
  author={Lee, Taeyoung},
  journal={IEEE Transactions on Automatic Control},
  volume={63},
  number={10},
  pages={3377--3392},
  year={2018},
  publisher={IEEE}
}

@inproceedings{srivatsan2017bingham,
  title={Bingham distribution-based linear filter for online pose estimation},
  author={Srivatsan, Rangaprasad Arun and Xu, Mengyun and Zevallos, Nicolas and Choset, Howie},
  booktitle={Robotics: Science and Systems},
  year={2017},
  organization={Robotics Science and Systems Foundation}
}

@article{pepe2024learning,
  title={Learning rotations},
  author={Pepe, Alberto and Lasenby, Joan and Chac{\'o}n, Pablo},
  journal={Mathematical Methods in the Applied Sciences},
  volume={47},
  number={3},
  pages={1204--1217},
  year={2024},
  publisher={Wiley Online Library}
}

@article{wang2024dual,
  title={Dual quaternion operations for rigid body motion and their application to the hand--eye calibration},
  author={Wang, Xiao and Sun, Haoxiang and Liu, Chenglin and Song, Hanwen},
  journal={Mechanism and Machine Theory},
  volume={193},
  pages={105566},
  year={2024},
  publisher={Elsevier}
}

@inproceedings{zhang2023revisiting,
  title={Revisiting rotation averaging: Uncertainties and robust losses},
  author={Zhang, Ganlin and Larsson, Viktor and Barath, Daniel},
  booktitle={Proceedings of the IEEE/CVF Conference on Computer Vision and Pattern Recognition},
  pages={17215--17224},
  year={2023}
}

@article{gerken2023geometric,
  title={Geometric deep learning and equivariant neural networks},
  author={Gerken, Jan E and Aronsson, Jimmy and Carlsson, Oscar and Linander, Hampus and Ohlsson, Fredrik and Petersson, Christoffer and Persson, Daniel},
  journal={Artificial Intelligence Review},
  volume={56},
  number={12},
  pages={14605--14662},
  year={2023},
  publisher={Springer}
}

@article{batzner20223,
  title={E (3)-equivariant graph neural networks for data-efficient and accurate interatomic potentials},
  author={Batzner, Simon and Musaelian, Albert and Sun, Lixin and Geiger, Mario and Mailoa, Jonathan P and Kornbluth, Mordechai and Molinari, Nicola and Smidt, Tess E and Kozinsky, Boris},
  journal={Nature communications},
  volume={13},
  number={1},
  pages={2453},
  year={2022},
  publisher={Nature Publishing Group UK London}
}
}

\end{document}